\documentstyle[12pt]{article}
\def\lq{\raise0.ex\hbox{$<$\kern-0.75em\raise-0.2ex\hbox{$_q$}}}
\newcommand{\be}{\begin{eqnarray}}
\newcommand{\ee}{\end{eqnarray}}
\begin{document}
\begin{titlepage}
\hspace*{12.0cm}HU-TFT-92-39

\hspace*{12.0cm}IFT-P/45-92
\vskip .50 cm

\begin{center}
{\Large Operator Formulation of $q$-Deformed Dual String Model}
\vskip 1.0 cm
M. Chaichian\\
{\it High Energy Physics Laboratory, Department of Physics,\\
University of Helsinki, Siltavuorenpenger 20 C,\\
SF-00170 Helsinki, Finland}
\vskip .50 cm
J.F. Gomes\\
{\it Instituto de F\'{i}sica Te\'{o}rica - UNESP,\\
Rua Pamplona 145,\\
01405 S\~{a}o Paulo SP., Brazil}
\vskip 0.5 cm
and
\vskip 0.5 cm
P. Kulish\\
{\it Steklov Mathematical Institute of the Academy\\
of Sciences, 191011 St. Petersburg, Russia}
\end{center}
 \vskip 2.0cm
\centerline {\bf Abstract}

We present an operator formulation of the $q$-deformed dual string
model amplitude
using an infinite set of $q$-harmonic oscillators. The formalism attains
the
crossing symmetry and factorization and allows to express the general
$n$-point function as a factorized product of vertices and propagators.
\end{titlepage}

The operator formalism \cite{fugove} has proven to be extremely useful in
formulating the dual amplitudes which have ultimately lead to the string
theory \cite{grscwi}. The main feature of this formalism is to provide
factoriza
   tion
of the dual amplitudes in such a way that they can be constructed, in the
tree-approximation, like Feynman diagrams out of the product of vertices and
propagators as the building blocks. The latters in turn possess the crossing
symmetry property in the sense that they are invariant under exchange of a
channel to its dual one, and so does the amplitude constructed out of them.

The factorization properties of the dual string model (for a review see,
e.g. \cite{alambeol}) have been shown by Fubini, Gordon and Veneziano (FGV)
\cite{fugove} in the operator formalism by introducing an infinite number of
har
   monic
oscillator operators $a^{(n)}_\mu$, one for each normal mode.

Due to the recent overwhelming interest in the quantum groups and the
$q$-deformation (see, e.g. \cite{fareta} for  refs.), in this letter we follow
the operator formulation proposed in \cite{fugove} to present a $q$-deformed
dua
   l
string amplitude by introducing an infinite set of $q$-deformed harmonic
oscillator operators ($q$-oscillators) \cite{mabichku} requiring the crossing
symmetry (duality) and factorization.

Attempts of similar types have been made by involving a finite
number of oscillators to obtain nonlinear trajectories \cite{coyuba}, or by
replacing \cite{ro} the ordinary $\Gamma$-functions appearing in the
Veneziano model by their $q$-analogs \cite{ex}.

We show that the constructed $q$-deformed dual model amplitudes using the
$q$-oscillators possess factorization and the crossing symmetry. Expressions
for the vertices and propagators are given. This allows us to construct the
deformed amplitudes as Feynman-like diagrams. As a simple illustration the
$q$-deformed 4-point amplitude is written down in the first order approximation
for the deformation parameter $q$.

Let us start by considering the ordinary (undeformed) 4-point dual amplitude
given by
\be
{\cal A}_4=\int^1_0\,dz\;z^{-\alpha(t)-1}(1-z)^{-\alpha(s)-1}\ ,
\ee
where crossing symmetry becomes manifest by the interchange of $z\rightarrow
1-z$. Here $\alpha(s)=\alpha's+\alpha_0$ is the linear Regge trajectory.
In order to factorize dual amplitudes such as (1) FGV \cite{fugove} introduced
an infinite set of oscillators
\be
[a^\mu_m\,,\,a^{\nu^+}_n]=\delta_{mn}g_{\mu\nu}\ ,
\ee
where $\mu,\nu$ are the four Lorentz indices; $m,n=1,2...$ . These oscillators
allow to write the term in the integrand of eq. (1) as \cite{fugove}
\be
(1-z)^{-A\bar{A}}=\prod\limits^\infty_{n=1}<0\,|\,e^{\frac{Aa_n}{\sqrt{n}}}\,
z^{na^+_na_n}\,e^{\frac{{\overline A}a^+_n}{\sqrt{n}}}| 0>\ ,
\ee
where in all the scalar products the contraction of the Lorentz indices of
$A,\,\bar{A},\,a_n$ and $a_n^+$ are understood. Thus from now on we omit
the Lorentz indices $\mu,\nu$ also in (2). Here the 4-vectors $A_\mu$ and
$\bar{A}_\mu$ can be taken to be any of the incoming or outgoing 4-momenta,
respectively.
The relation (3) can be easily verified by considering the orthonormal
basis for the Fock space of the states built from the vacuum
$|0>,\;a_n|0>=0$,
as
\be
|\lambda >\equiv |\lambda_1,\lambda_2,...,\lambda_i,...>
=\prod\limits^{\infty}_{i=1}\;
\frac{(a^+_i)^{\lambda_i}}{\sqrt{\lambda_i!}}|0>\ ,
\ee
with $<\lambda|\lambda'>=
\delta_{\lambda\lambda'}\equiv\delta_{\lambda_1\lambda'_1}
\delta_{\lambda_2\lambda'_2}\cdots\ .$

In order to perform the $q$-deformation within the operator formalism, it
seems natural to employ the $q$-oscillators and their corresponding Fock
space in the r.h.s. of eq. (3). For this purpose let us consider an
infinite
set of $q$-oscillators \cite{mabichku} characterized by a deformation parameter
$q$ and satisfying the following commutation relations:
\be
a_m a^+_m -qa^+_m a_m=q^{-N_m},\hspace{3.5cm}\nonumber\\
\ [N_m,a^+_m ]=a^+_m,[N_m,a_m]=-a_m \hspace{2.2cm}\nonumber\\
\ [a_n,a_m ]=[a^+_n,a^+_m]=[a^+_n,a_m]=0\ ,\ m\neq n\ ,
\ee
where $N_m$ is the $q$-number operator corresponding to the mode $m$. These
operators also build up the corresponding Fock space of states with the
orthonormal basis by successive application of $a^+_n$ to the vacuum
$|0>_q,\ a_n|0>_q=0$, as
\be
\bigl|\lambda>_q\equiv |\lambda_1,\lambda_2,...,\lambda_i,...>_q
=\prod\limits_i
\frac{(a^+_i)^{\lambda_i}}{\sqrt{[\lambda_i]!}}\bigr|0>_q\ ,
\ee
with $\lq\;\ \lambda|\lambda'>_q=\delta_{\lambda\lambda'}\equiv
\delta_{\lambda_1\lambda'_1}\delta_{\lambda_2\lambda'_2}\cdots\ ,$ and
\be
N_m|\lambda>_q=\lambda_m|\lambda>_q\ ,\ a^+_ma_m|\lambda>_q
=[\lambda_m]|\lambda>_q\ ,
\ee
where $[x]\equiv\frac{q^x-q^{-x}}{q-q^{-1}}\ ,\ [n]!=[1][2]\cdots[n]$
and $[0]!=1$.
\vskip 0.5 cm
For the $q$-deformed version of eq. (3) we propose to replace in the r.h.s.
of (3) the oscillators $a_n,a^+_n$ by their $q$-deformed counterparts given
by (5). Furthermore, guided by the normalization factors
$\frac{1}{\sqrt{[\lambda_i]!}}$ for the state vectors in (6), we also replace
th
   e exponent
$e^x$ by the $q$-exponential function $e^x_q\equiv\sum\limits^\infty_{n=0}
\frac{x^n}{[n]!}$ . Then it is easy to show the following identity
\be
F(A,\bar{A},z)\equiv\prod\limits^\infty_{n=1}\sum\limits^\infty_{\ell=0}
\left(\frac{A\bar{A}}{n}\right)^\ell\frac{z^{n[\ell]}}{[\ell]!}
=\prod\limits^\infty_{n=1}\lq\ \ 0\,| e_q^{\frac{Aa_n}{\sqrt{n}}}\,
z^{na^+_na_n}\,e_q^{\frac{{\overline A}a^+_n}{\sqrt{n}}}\,|0>_q\ .
\ee
Eq. (8) yields (3) for the undeformed case as the limit $q\rightarrow 1$.
As pointed out earlier, the symmetry $z\rightarrow 1-z$ in (1) was
essential for
having the duality in the ordinary case. However, due to the deformation
carried out in (8), $F(A,\bar{A},z)$ differs now from
$(1-z)^{A\bar{A}}$ and
thus we are unambiguously lead to deform also the other term
$z^{-\alpha_t-1}$
in (1) in a similar manner as in (8), i.e.
\be
D(B,\bar{B},z)\equiv\prod\limits^\infty_{m=1}\ \sum\limits^\infty_{k=0}
\left(\frac{B\bar{B}}{m}\right)^k\frac{(1-z)^{m[k]}}{[k]!}
=\prod\limits^\infty_{m=1}\ \lq\;\  0|\,e_q^{\frac{Bb_m}{\sqrt{m}}}
(1-z)^{mb^+_mb_m}\,e_q^{\frac{{\overline B}b^+_m}{\sqrt{m}}}|\,0>_q\ ,
\ee
where we have introduced a new independent set of $q$-oscillators $b_m,b^+_m$
which satisfy the commutation relations (5) and commute with $a_n,a^+_n$.
The 4-Lorentz vectors $B_\mu$ and $\bar{B}_\mu$ are again related to incoming
and outgoing momenta the scalar product of which gives in the case of
four-point amplitude the $t$-channel (dual to $s$-channel) Regge trajectory,
and is given by $\alpha(t)+1$. The $q$-deformed 4-point amplitude can thus
be written in a manifestly crossing symmetric way as
\be
{\cal A}^q_4=\int^1_0dz D(B,\bar{B},z)F(A,\bar{A},z)\ ,
\ee
with $F$ and $D$ given by (8) and (9). Notice that the crossing is obtained
by the interchange $A,\bar{A}\rightarrow B,\bar{B}$.

In terms of $q$-oscillators $a_n,a^+_n,b_n,b^+_n$, as introduced in (8) and
(9), the 4-point $q$-amplitude can be expressed as
\be
{\cal A}^q_4=\int^{1}_{0}\,dz\prod\limits^{\infty}_{m,n=1}\;\lq\;\  0,0|
e_q^{\frac{Bb_m}{\sqrt{m}}}\,e_q^{\frac{Aa_n}{\sqrt{n}}}
(1-z)^{mb^+_mb_m}z^{na^+_na_n}\,e_q^{\frac{{\overline B}b^+_m}{\sqrt{m}}}
e_q^{\frac{{\overline A}a^+_n}{\sqrt{n}}}|0,0>_q\ .
\ee
Since the $q$-oscillators $(a_n,a^+_n)$ and $(b_n,b^+_n)$ are independent,
the vacuum state $|0,0>_q$ is a direct product $|0>_{a,q}\otimes |0>_{b,q}$,
out of which the complete set of orthonormal states is built up:
\be
\sum\limits_{\lambda^a,\lambda^b}|\lambda^a,\lambda^b>_q\,\lq\
\lambda^a,\lambda^b
|=\sum\limits_{\lambda^a,\lambda^b}|\lambda^a>_q\otimes|
\lambda^b>_q\,\lq\;\ \lambda^a|\otimes\,\lq\ \lambda^b|=1\ .
\ee
Inserting the identity (12) twice into (11) and using (6) and (7), we can
rewrite the $q$-amplitude (11) in a factorized form
\be
{\cal A}^q_4=\sum\limits_\lambda
V(A,B;\lambda)\,D\,(\lambda;A,\bar{A},B,\bar{B}
   )
V(\bar{A},\bar{B};\lambda)^{+}\ ,
\ee
where $\lambda$ denotes both sets of $\lambda^a$ and $\lambda^b$, and the
$q$-vertices are defined as
\be
V(A,B;\lambda)=\lq\;\ 0,0\,|\prod\limits^\infty_{m,n=1}
e_q^{\frac{Aa_n}{\sqrt{n}}}\,
e_q^{\frac{Bb_m}{\sqrt{m}}}|\lambda_a,\lambda_b>_q\ ,
\ee
and the $q$-propagator as
\be
D(\lambda;A,\bar{A},B,\bar{B})=\int^1_0dz\,\lq\;\ \lambda_a,\lambda_b
\left|(1-z)^{R_a}
z^{R_b}\right|\lambda_a,\lambda_b>_q\ ,
\ee
with
\be
R_a=\sum\limits^\infty_{n=1}n\,a^+_na_n\ ,\ R_b=\sum\limits^{\infty}_{m=1}
mb^+_mb_m\ .
\ee
We should point out that from the example of the 4-point amplitude (13) we
can derive the vertex function (14) which has only one excited leg
characterized by the intermediate states $|\lambda_a,\lambda_b>$. However, in
order to write down a general multipoint amplitude as a product of vertices
and propagators, we need to define in addition a new vertex function $\Gamma$
which has two excited legs. This new vertex can be extracted from a
$q$-deformed 5-point amplitude ${\cal A}^q_5$ which can be written down in a
factorized form. For this purpose we proceed exactly as in the previous
case of 4-point function. Let us first consider the ordinary undeformed
5-point amplitude, which in terms of undeformed oscillators is given by
[1]
\be
{\cal A}_5=\int^1_0du_1\int^1_0du_2\,u^{-\alpha(s_1)-1}_1\,
u^{-\alpha(s_2)-1}_2<0|G|0>\ ,
\ee
where
\be
G=\prod\limits^{\infty}_{n=1}e^{\frac{Aa_n}{\sqrt{n}}}
u^{R_a}_2\,e^{\frac{\bar{p}a^+_n}{\sqrt{n}}}\ \
e^{\frac{pa_n}{\sqrt{n}}}\,u^{R_a}_1\,e^{\frac{{\overline A}a^+_n}
{\sqrt{n}}}\ ,
\ee
and $p=\bar{p}$ is the momentum of the unexcited leg in the vertex.

As it occured in the previous case of ${\cal A}^q_4$, in order to preserve
duality a new set of deformed oscillators had to be introduced. Here we
need two additional sets of independent $q$-oscillators $b^{(i)}_n,\,
b^{(i)^+}_n,\; i=1,2$, one for each integration variable $u_1$ and $u_2$.
We thus can  write the $q$-deformed 5-point amplitude as
\be
{\cal A}^q_5=\int^1_0du_1\int^1_0du_2\,\lq\;\  0,0,0\left| U_1U_2G^q\right|0,
0,0>_q\ ,
\ee
where
\be
U_i=\prod\limits^{\infty}_{m=1}e^{\frac{B^{(i)}b^{(i)}_m}
{\sqrt{m}}}_q\,(1-u_i)^{mb^{(i)^+}_mb_m^{(i)}}
e^{\frac{{\overline B}^{(i)}b_m^{(i)}}{\sqrt{m}}}_q\ \ ,\ \ i=1,2,
\ee
and $B_i,\bar{B}_i$ correspond to different momenta of the external
particles with their scalar products producing the Mandelstam variables $s_1$
and $s_2$ of the corresponding channels, and
\be
G^q=\prod\limits^{\infty}_{n=1}e^{\frac{Aa_n}{\sqrt{n}}}_qu^{R_a}_2\,e_q
^{\frac{{\overline p}a^+_n}{\sqrt{n}}}e_q^{\frac{pa_n}{\sqrt{n}}}\,u_1^{R_a}
e^{\frac{{\overline A}a^+_n}{\sqrt{n}}}_q\ ,
\ee
where $a_n,a^+_n$ are the $q$-oscillators (5).\\
\noindent  Inserting now four times the
unity operator in terms of complete sets of orthonormal states,
\be
\sum\limits_{\lambda}|\lambda>_q\,\lq\;\ \lambda|\equiv\sum\limits_{\lambda^a,
\lambda^{b_1},\lambda^{b_2}}|\lambda^a,\lambda^{b_1},\lambda^{b_2}>_q\,
\lq\;\ \lambda^a,\lambda^{b_1},\lambda^{b_2}|=1\ ,
\ee
into (19), we arrive at the factorized form for the $q$-deformed 5-point
amplitude ${\cal A}^q_5$:
\be
{\cal A}^q_5=\sum\limits_{\lambda,\lambda'}V(A,B_1;\lambda)D(\lambda;A,\bar{A},
B_1,\bar{B}_1)\Gamma(\lambda,\lambda';p,\bar{p},\bar{B}_i)\nonumber\\
D(\lambda';A,\bar{A},B_2,\bar{B}_2)V(\bar{A},B_2;\lambda')^{+}\ ,
\ee
where the new vertex $\Gamma$ with two excited legs is given by
\be
\Gamma(\lambda,\lambda';p,\bar{p},\bar{B}_i)=\lq\;\
\lambda|\prod\limits^{\infty
   }
_{m,n,\ell=1}e^{\frac{{\overline p}a^+_n}{\sqrt{n}}}_q
e^{\frac{{\overline B}_1b^{(1)^+}_m}
{\sqrt{m}}}_qe^{\frac{pa_n}{\sqrt{n}}}_q
e^{\frac{{\overline B}_2b^{(2)}_\ell}{\sqrt{\ell}}}_q
|\lambda'>_q\ \ ,\ i=1,2\ .
\ee
We notice that in the undeformed limit $q\rightarrow 1$, from (20) one
obtains
\be
<0_i|U^i|0_i>=\prod\limits^{\infty}_{m=1}e^{\frac{B_i{\overline B}_i(1-u_i)
^m}{m}}=u^{-B_i{\overline B}_i}_i\ ,i=1,2\ ,
\ee
whilest the remaining factor in (19), after using the usual
Campbell-Baker-Hausdorff relation, becomes
\be
G=\exp\{\sum\limits^{\infty}_{n=1}\frac{1}{n}(\bar{A}A(u_1u_2)^n+\bar{A}p
u^n_1+\bar{p}Au^n_2)\}\ ,
\ee
coinciding with the corresponding expression in [1].

Let us consider as an illustration the example of $q$-deformed 4-point
amplitude (10) in the approximation $q=e^\epsilon=1+\epsilon+
\frac{\epsilon^2}{2!}+0(\epsilon^3)$. Using the expansions $[\ell]=\ell
(1+\frac{\epsilon^2}{6}(\ell^2-1))+0(\epsilon^4)$ and $[\ell]!=\ell!
(1+\frac{\epsilon^2}{36}\ell(\ell-1)(2\ell+5))+0(\epsilon^4)$, we obtain
\be
\frac{x^{n{[\ell]}}}{[\ell]!} = \frac{x^{n\ell}}{\ell!}(1+
\frac{\epsilon^2}{6}n(\ell+1)
\ell(\ell-1)\ln\,x-\frac{\epsilon^2}{36}\ell(\ell-1)(2\ell+5))+0(\epsilon^4)\ .
\ee
Using (27) with $x=z$ and $1-z$ in (8) and (9), respectively, and
employing the following identities
\be
\sum\limits^{\infty}_{r=0}\frac{x^r}{r!}(r+1)r(r-1)=x^2(x+3)e^x\
,\hspace{0.5cm}
\nonumber\\
\sum\limits^{\infty}_{r=0}\frac{x^r}{r!}\,r(r-1)(2r+5)=(2x^3+9x^2)e^x\ ,
\nonumber
\ee
we finally arrive at
\be
{\cal A}_4=B(-\alpha(s),-\alpha(t))-\frac{\epsilon^2}{3!}\Delta\
,\hspace{2.0cm}
\ee
where
\be
\Delta=\frac{1}{3!}\sum\limits^{\infty}_{n=1}\frac{(\alpha(s)+1)^2}
{n^2}\{9B(2n-\alpha(s),-\alpha(t))\hspace{1.0cm}\nonumber\\
+2\frac{\alpha(s)+1}{n}B(2n-\alpha(s),-\alpha(t))\}\nonumber\\
+\sum\limits^{\infty}_{m,n=1}\frac{(\alpha(s)+1)^2}{mn}\{3B(2m-\alpha(s),
n-\alpha(t))\hspace{1.0cm}\nonumber\\
-\frac{\alpha(s)+1}{m}B(3m-\alpha(s),n-\alpha(t))\}+(s\leftrightarrow t)\ ,
\ee
$B(-\alpha(s),-\alpha(t))$ is the usual beta-function, and
$A\bar{A}=\alpha(s)+1
,\ B\bar{B}=\alpha(t)+1$ have been used.

To conclude we have proposed the $q$-deformed four- and five-point
amplitudes in a manifestly dual and factorized form by using $q$-oscillators
in the operator formulation of dual string model. This method has provided
us with all the ingredients needed to build up a general $n$-point
$q$-amplitude as a product of vertices and propagators in a Feynman-like
diagram manner. The advantage of using the operator formulation in the
case of $q$-deformed theory, in addition to providing the explicit
factorization, is that algebraic considerations such as conditions for
absence of unphysical ghost states (see, e.g. [3]), may be treated
in a similar way as in the ordinary undeformed case. Further considerations
in the properties of the proposed $q$-deformed dual model are under
study.
\vskip 1.0 cm

We would like to thank Claus Montonen for useful discussions. One of
us (J.F.G.) gratefully acknowledges the hospitality of the Research
Institute for Theoretical Physics, University of Helsinki, and
FUNDUNESP (Brasil) for partial financial support.

\end{document}